\documentclass[sigconf]{acmart}
\usepackage{enumitem}
\usepackage{algorithm}
\usepackage{algpseudocode}

\usepackage{amsmath}      
\usepackage{color}
\usepackage{colortbl}
\usepackage{multirow}
\usepackage{multicol}
\definecolor{lightblue}{RGB}{235,245,255}
\definecolor{mygreen}{HTML}{228B22}

\let\oldhat\hat

\renewcommand{\hat}[1]{\oldhat{\mathbf{#1}}}

\AtBeginDocument{%
  }

\setcopyright{acmlicensed}
\copyrightyear{2018}
\acmYear{2018}
\acmDOI{XXXXXXX.XXXXXXX}
\acmConference[Conference acronym 'XX]{Make sure to enter the correct
  conference title from your rights confirmation email}{June 03--05,
  2018}{Woodstock, NY}
\acmISBN{978-1-4503-XXXX-X/2018/06}

\begin{document}

\title{RankGR: Rank-Enhanced Generative Retrieval with Listwise Direct Preference Optimization in Recommendation}

\author{Kairui Fu}
\authornote{Equal contribution}
\affiliation{%
  \institution{Zhejiang University}
  \city{Hangzhou}
  \country{China}}
\email{fukairui.fkr@zju.edu.cn}

\author{Changfa Wu}
\authornotemark[1]
\affiliation{%
  \institution{Alibaba Group}
  \city{Beijing}
  \country{China}}
\email{wuchangfa.wcf@alibaba-inc.com}

\author{Kun Yuan}
\affiliation{%
  \institution{Alibaba Group}
  \city{Beijing}
  \country{China}}
\email{yuankun.yuan@taobao.com}

\author{Binbin Cao}
\affiliation{%
  \institution{Alibaba Group}
  \city{Beijing}
  \country{China}}
\email{simon.cbb@taobao.com}

\author{Dunxian Huang}
\affiliation{%
\institution{Alibaba Group}
\city{Hangzhou}
\country{China}
}
\email{dunxian.hdx@alibaba-inc.com}

\author{Yuliang Yan}
\affiliation{%
\institution{Alibaba Group}
\city{Hangzhou}
\country{China}
}
\email{yuliang.yyl@alibaba-inc.com}

\author{Junjun Zheng}
\affiliation{%
\institution{Alibaba Group}
\city{Hangzhou}
\country{China}
}
\email{fangcheng.zjj@alibaba-inc.com}

\author{Jianning Zhang}
\affiliation{%
\institution{Alibaba Group}
\city{Hangzhou}
\country{China}
}
\email{zhangjianning.zjn@alibaba-inc.com}

\author{Silu Zhou}
\affiliation{%
\institution{Alibaba Group}
\city{Hangzhou}
\country{China}
}
\email{silu.zsl@alibaba-inc.com}

\author{Jian Wu}
\affiliation{%
\institution{Alibaba Group}
\city{Beijing}
\country{China}
}
\email{joshuawu.wujian@alibaba-inc.com}

\author{Kun Kuang}
\affiliation{%
\institution{Zhejiang University}
\city{Hangzhou}
\country{China}
}
\email{kunkuang@zju.edu.cn}


\renewcommand{\shortauthors}{K Fu et al.}

\begin{abstract}
  Generative retrieval (GR) has emerged as a promising paradigm in recommendation systems by autoregressively decoding identifiers of target items. Despite its potential, current approaches typically rely on the next-token prediction schema, which treats each token of the next interacted items as the sole target. This narrow focus 1) limits their ability to capture the nuanced structure of user preferences, and 2) overlooks the deep interaction between decoded identifiers and user behavior sequences.
  In response to these challenges, we propose \textbf{RankGR}, a \textbf{Rank}-enhanced \textbf{G}enerative \textbf{R}etrieval method that incorporates listwise direct preference optimization for recommendation. RankGR decomposes the retrieval process into two complementary stages: the \textbf{I}nitial \textbf{A}ssessment \textbf{P}hase (IAP) and the \textbf{R}efined \textbf{S}coring \textbf{P}hase (RSP). In IAP, we incorporate a novel listwise direct preference optimization strategy into GR, thus facilitating a more comprehensive understanding of the hierarchical user preferences and more effective partial-order modeling. The RSP then refines the top-$\lambda$ candidates generated by IAP with interactions towards input sequences using a lightweight scoring module, leading to more precise candidate evaluation.
  Both phases are jointly optimized under a unified GR model, ensuring consistency and efficiency.
  Additionally, we implement several practical improvements in training and deployment, ultimately achieving a real-time system capable of handling nearly ten thousand requests per second.
  Extensive offline performance on both research and industrial datasets, as well as the online gains on the "Guess You Like" section of Taobao, validate the effectiveness and scalability of RankGR.
\end{abstract}




\maketitle

\section{Introduction}

\begin{figure}[htb]
    \centering
    \vspace{-0.15cm}
    \includegraphics[width=1.0\linewidth]{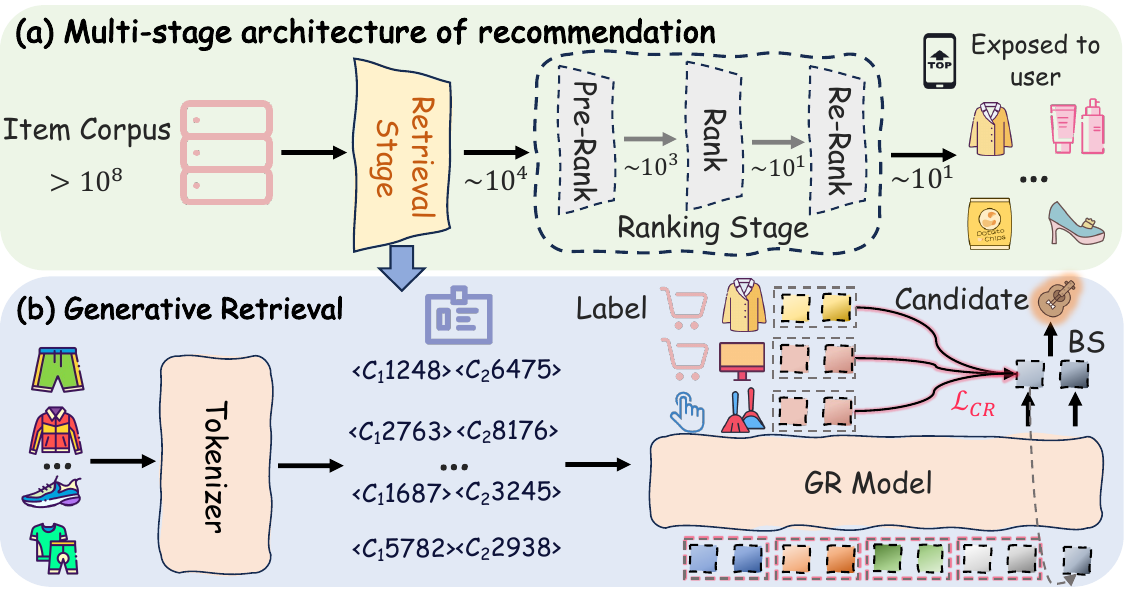}
    \vspace{-0.4cm}
    \caption{(a) Multi-stage architecture in modern recommender systems. (b) A brief diagram of generative retrieval.}
    \label{fig: Intro}
    \vspace{-0.17cm}
\end{figure}

Recommender systems play a crucial role in uncovering latent user interests and mitigating the growing challenge of information overload~\cite{feng2021news}. This benefit has led to widespread adoption across diverse domains, including e-commerce~\cite{liang2025tbgrecall,wang2026pi2i} and social media~\cite{zhu2025rankmixer,zhang2025onetrans}. To manage the massive scale of users and items, existing approaches typically divide the process into two stages: retrieval and ranking. As illustrated in Figure~\ref{fig: Intro}(a), the retrieval stage rapidly filters out thousands of potentially relevant items from a massive corpus containing billions, which are then passed to the ranking stage for more precise evaluation. Consequently, the retrieval stage presents a key challenge in building scalable recommender systems to balance response time, coverage, and relevance~\cite{wang2025beyond}.

Recent advances in generative models~\cite{brown2020language,yang2025qwen3} have sparked interest in generative retrieval (GR) as a promising alternative to traditional methods like collaborative filtering~\cite{yang2020large,he2017neural} or two-tower architectures~\cite{kang2018self,li2019multi,wang2025beyond}. Unlike these methods, which typically rely on similarity matching between user and item embeddings, generative retrieval in Figure~\ref{fig: Intro}(b) aims to directly produce a set of candidate items based on user behavior sequences, thereby offering new opportunities for capturing sequential patterns in an end-to-end manner. TIGER~\cite{rajput2023recommender} pioneers this paradigm by introducing semantic identifiers (SIDs) using several tokens derived from quantized semantic embeddings of item content features, enabling the use of Transformer-based sequence-to-sequence models to autoregressively decode the SID of the next item. Subsequent work FORGE~\cite{fu2025forge} improves performance by optimizing SIDs using collaborative signals and reducing item collisions in industrial settings. COBRA~\cite{yang2025sparse} combines the strengths of sparse and dense representations through a cascading architecture, effectively mitigating the information loss typically introduced during quantization.

Despite the successful deployment of these methods in real-world platforms~\cite{fu2025forge,deng2025onerec,zheng2025ega}, their prevalent use of \textit{next token prediction} (NTP) may not be well-suited for generative retrieval due to the two fundamental challenges:
\romannumeral1) \textbf{Insufficient modelling of the partial order of user preference}. In industrial recommender systems, training data for each historical sequence is typically derived from user feedback on exposed samples within a session.
However, conventional NTP operates at the token level and treats these signals as \textit{isolated point-wise examples}, which lack explicit modeling at the item level and fail to capture the underlying partial order preference among items. For example, within a single session, user preference for the \textit{coat} and \textit{computer} purchased is substantially stronger than the preference for the clicked \textit{broom} or the \textit{chips} that were exposed but uninteracted
(i.e., \raisebox{-0.02in}{\includegraphics[height=0.13in, width=0.15in]{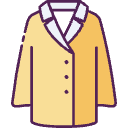}} \textbf{=} \raisebox{-0.02in}{\includegraphics[height=0.11in, width=0.13in]{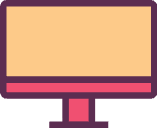}} \textbf{>} \raisebox{-0.02in}{\includegraphics[height=0.12in, width=0.15in]{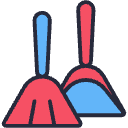}} \textbf{>} \raisebox{-0.04in}{\includegraphics[height=0.16in, width=0.17in]{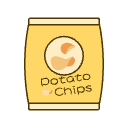}}).
The inability to capture such nuanced preference hierarchies hinders the ability to distinguish the truly relevant items, ultimately leading to suboptimal recommendations. 
\romannumeral2) \textbf{Lack of cross-effects between the candidates and user behavior sequences}. During inference, GR models predict the next candidate tokens (e.g., <$C_1$1687>) through the inner product of a single hidden state with the entire vocabulary, which resembles a two-tower matching framework. In recommender systems, this approximation only captures shallow interactions and overlooks the rich relationships between each candidate and the complete behavior sequence. As a result, GRs fall short in effectively modeling the evolving user-item space, limiting their performance in practical scenarios.

To address these challenges, this paper presents \textbf{RankGR}, a novel framework designed for \textbf{Rank}-enhanced \textbf{G}enerative \textbf{R}etrieval in recommendation. 
The core idea of our approach is to explicitly model a deep understanding of the partial order of user preferences during GR training while enabling meaningful interactions between candidate items and the behavior sequence at an affordable computational cost.
With this principle, RankGR integrates two key components: \textbf{\romannumeral1)} The \textbf{Initial Assessment Phase} (IAP) goes beyond the point-wise modeling of conventional NTP by introducing a novel training schema named listwise direct preference optimization (LDPO), which is specifically tailored to capture the hierarchical and partially ordered nature of user preferences inherent in recommendation. Based on the standard GR model, we start by modeling the overall probability of each item within a session with the combination of its corresponding tokens. During training, explicit reward signals in recommendations (e.g., purchases, clicks) are incorporated via supervised contrastive learning to guide the model in understanding the relative quality of items within a session. This strategy eliminates the need for an explicit reward model, thus significantly simplifying the training process. \textbf{\romannumeral2)} Subsequently, the \textbf{Refined Scoring Phase} (RSP) builds upon the hidden states generated by IAP and the top-$\lambda$ decoded candidate SIDs. It employs a lightweight scoring module to enable deep interaction between each candidate and the rich semantic information embedded in the input sequence, leading to fine-grained scoring for each candidate. RSP breaks the limitations of inner-product-based approximations commonly used in IAP and existing methods, resulting in more accurate and semantically informed scoring of candidate SIDs. 
The lightweight module within RSP is jointly trained with IAP, thus forming an efficient training framework.
Furthermore, to 
handle tens of thousands of requests per second in industrial settings, we also propose several engineering optimizations that guarantee real-time and high-fidelity responsiveness to user requests. 
In summary, the main contributions of our paper can be summarized as follows:
\begin{itemize}[leftmargin=*]
    \item We investigate the fundamental limitations of current GRs, particularly their inability to capture hierarchy preference and rich interactions between retrieved items and user behavior sequences, which hinder their ability to preserve several interesting items for the subsequent ranking stage.
    \item We propose RankGR, a rank-enhanced generative retrieval framework for recommendation. Equipped with the listwise direct preference optimization in IAP and the extra scoring module in RSP, RankGR breaks the limitations of existing methods and provides new directions for the further development of GR.
    \item Extensive offline comparison and additional analysis demonstrate the effectiveness of RankGR. Specifically, deployed in the Guess You Like Section for Taobao's homepage recommendation, RankGR brings a notable 1.08\% increase in Item Page Views.
\end{itemize}

\section{Related Work}

\subsection{Generative Retrieval}

The development of generative models has led to the development of generative retrieval (GR) in recommender systems, which enables more flexible modeling by autoregressively generating item representations from historical behaviors. Representative methods like HSTU~\cite{zhai2024actions} adopt stacked hierarchical sequential transduction units to enhance feature interactions and selection, demonstrating how GR can benefit from scaling laws observed in large-scale recommendation systems. To prevent information loss due to the lack of explicit feature crosses, MTGR~\cite{han2025mtgr} incorporates a feature structure that aligns with traditional recommenders, ensuring compatibility within the generative framework. TIGER~\cite{rajput2023recommender} leverages the capability of generative models~\cite{yang2025qwen3} to construct semantic identifiers and employs beam search to retrieve the next potential items. FORGE~\cite{fu2025forge} then introduces improvements in both the generation and post-processing of SIDs, achieving notable performance gains with the same model architecture. ETEGRec~\cite{liu2025generative}, on the other hand, integrates the generation of semantic IDs with GR training in an end-to-end manner, enabling more coherent learning of user preferences. Further methods ~\cite{deng2025onerec, chen2025onesearch} advance this line of research by incorporating reinforcement learning to better align with user preferences, thus improving long-term engagement and satisfaction.

\subsection{Preference Alignment}

Preference alignment is a crucial aspect of training large language models (LLMs) to ensure their outputs align with human expectations. Several methods~\cite{chowdhury2024provably,rafailov2023direct,ouyang2022training,zhong2024dpo} have been proposed in this area to offer distinct approaches to integrating human feedback into the learning process. One prominent method is Direct Preference Optimization (DPO), which directly incorporates human preferences into the model’s training by optimizing a loss function based on pairwise comparisons~\cite{rafailov2023direct}. Subsequent Multi-sample Direct Preference Optimization (mDPO) extends DPO by focusing on group-wise characteristics, which is more effective in optimizing collective characteristics for generative models than single-sample comparison~\cite{wang2024preference}. Another approach, Preference Flow Matching (PFM), uses flow-based generative models to transform less preferred data into preferred outcomes, reducing reliance on extensive fine-tuning~\cite{kim2024preference}. In recommendation, several approaches~\cite{deng2025onerec,chen2025onesearch} also utilize DPO to capture fine-grained user preferences. The fundamental difference between RankGR and these works in recommender systems lies in the capture of multi-level preferences and the explicit interaction between candidate items and historical sequence.

\section{Methodology}

\begin{figure*}[htb]
    \centering
    \vspace{-0.3cm}
    \includegraphics[width=0.83\linewidth]{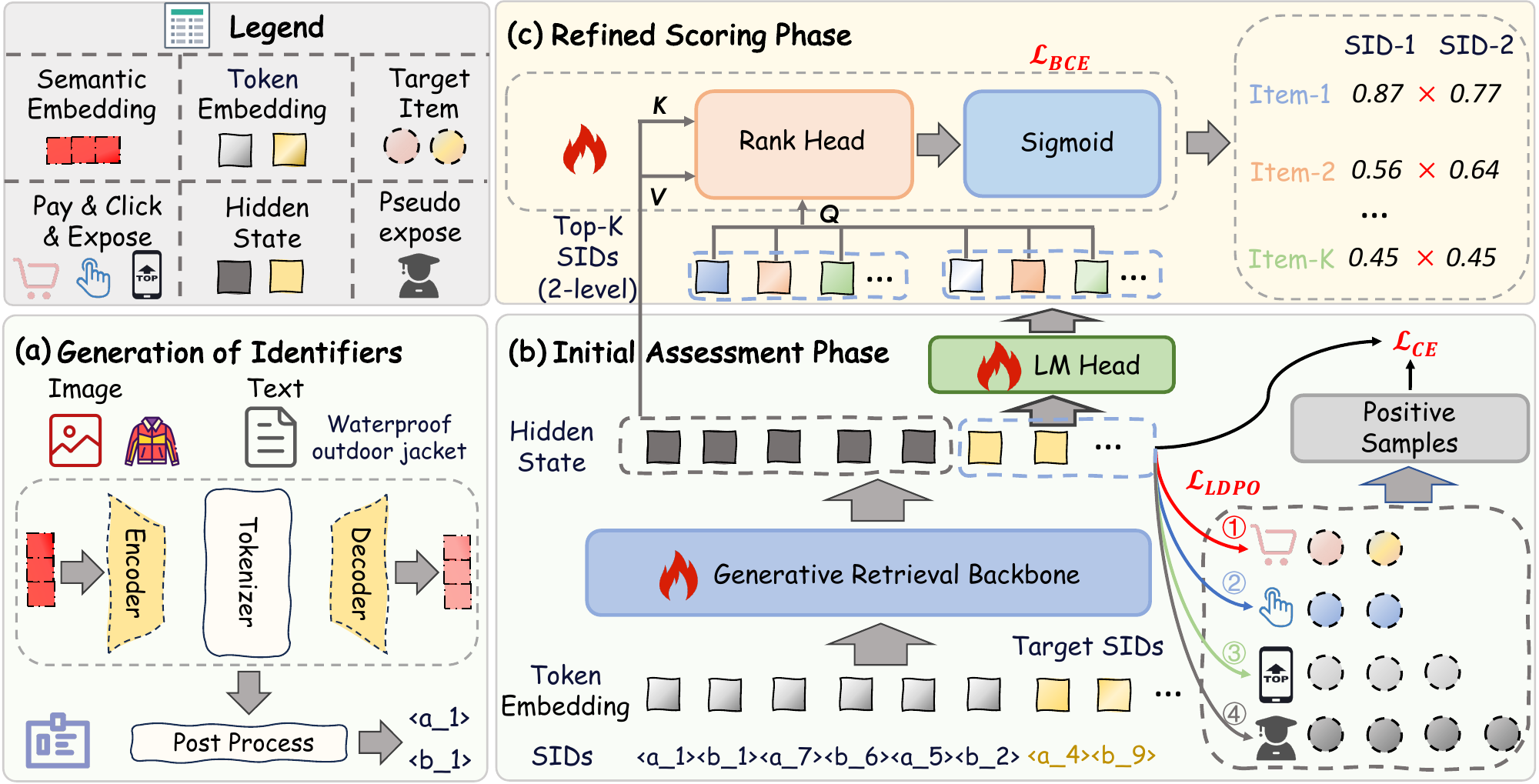}
    \caption{Overview of the training of RankGR. (a) The generation process of semantic identifiers. (b) The initial assessment phase aims to capture both the sequential pattern and the partial order relations. (c) The refined scoring phase to provide a more accurate prediction of those candidates from the initial assesment.}
    \label{fig: Method}
    \vspace{-0.4cm}
\end{figure*}

In this section, we will describe the proposed \textbf{Rank}-enhanced \textbf{G}enerative
\textbf{R}etrieval (RankGR) framework in detail. Specifically, Section~\ref{preliminary} presents a brief overview of the workflow for both SID generation and GR training as shown in Figure~\ref{fig: Method}(a). Subsequent Section~\ref{IAP} and Section~\ref{RSP} detail the Initial Assesment Phase (IAP) with listwise direct preference optimization and the Refined Scoring Phase (RSP) for more precise assesment in order. We further elaborate on the overall deployment pipeline with several engineering optimizations in Figure~\ref{fig:deployment} of  Section~\ref{deployment}.

\subsection{Problem Formulation}
\label{preliminary}

In the context of recommender systems, there exists a set of users $\mathcal{U}=\{u_1,u_2,\cdots,u_{N_u}\}$ and items $\mathcal{I}=\{i_1,i_2,\cdots, i_{N_i}\}$, with their respective cardinalities denoted as $ N_u $ and $ N_i $, respectively. 
The goal of generative retrieval (GR) is to autoregressively decode the identifier of the next item based on the identifiers of the historical interactions. Unlike previous research~\cite{zhai2024actions,liang2025tbgrecall}, which typically encodes each item using meaningless single-token identifiers, the advancement of generative models~\cite{team2024qwen2, Qwen-VL} enables GR to encode each item $i$ using rich multi-modal features $\mathcal{H}^i$.

As illustrated in Figure~\ref{fig: Method} (a), the features of item $i$ are encoded into the semantic identifier (SID) composed of discriminative multi-level semantic codewords $\{c^i_1, \dots, c^i_m\}$\footnote{In this paper, we use superscripts and subscripts to denote the 1st level (e.g., each item) and 2nd level classification (e.g., the codeword of items), separately.} through the residual quantized variational autoencoder (RQ-VAE)~\cite{zeghidour2021soundstream}, where $ m $ denotes the number of codewords within SID. Formally, the encoding process of the $l$-th codeword $c^i_l$ can be formulated as:
\begin{equation}
    c^i_l = \arg\min_{c \in [1, m]} \left\| \mathcal{H}^{i}_l - q_l^c \right\|,
\end{equation}
where $\mathcal{H}^{i}_l$ is obtained as the residual between the input feature $\mathcal{H}^{i}_{l-1}$ and the nearest vector $q_{l-1}^c$ from the previous codebook. At the first level ($l=1$), the initial residual is defined as $\mathcal{H}^{i}_{1}=\mathcal{H}^{i}$.

Once the SIDs are allocated, they will be fixed, and GR can be trained over the entire dataset using the next-token prediction loss:
\begin{equation}
\label{eq:autoregressive_decoding}
    \mathcal{L}_{\text{NTP}} = \frac{1}{N_u} \sum_{j=1}^{N_u} \sum_{t=1}^{T} \sum_{k=1}^{m} -\log P_{\theta, \tau}(c^{i(t)}_k \mid x^j_{<t}, c^{i(t)}_{<k}),
\end{equation}
where $T$ denotes the length of interaction for user $u(j)$ and $i(t)$ is the $t$-th interacted item. $x^j_{<t}$ represents the concatenation of SIDs from the historical interactions of user $j$, and each $x^j_t$ is made of an $m$-level SID $[c^{i(t)}_1, c^{i(t)}_2, \dots, c^{i(t)}_m]$. Besides, $\theta$ and $\tau$ denote the parameters of the GR backbone model $\mathcal{M}_{\text{GR}}$ and the tokenizer for codewords, respectively.

\subsection{Initial Assesment Phase}
\label{IAP}

As described in Equation~\ref{eq:autoregressive_decoding}, the current training paradigm typically operates at the token and codeword level, thereby failing to explicitly model the intensity of user preference for items composed of exactly $m$ codewords. 
To address this, we propose the Initial Assesment Phase (IAP) based on the current GR schema in Equation~\ref{eq:autoregressive_decoding} to capture the hierarchical structure of user preferences over items.

\subsubsection{Item level modeling.} Given the fact that each item $i(t)$ is represented by a sequence of $m$ codewords $\{c^{i(t)}_1, \dots, c^{i(t)}_m\}$, the GR model would autoregressively generate a sequence of the corresponding hidden states $\{h^{i(t)}_1, h^{i(t)}_2, \dots, h^{i(t)}_m\}$ for the next item prediction: 
\begin{equation}
\label{hiden_status}
    h^{i(t)}_l = \mathcal{M}_{\text{GR}}(x^j_{<t}, c^{i(t)}_{<l}), \forall l \in [1, m].
\end{equation}
The hidden states are then used to compute the similarity across all codewords in this level. Specifically, the similarity between the hidden state $h^{i(t)}_l$ and the corresponding codeword $c^{i(t)}_l$ is computed via the inner product followed by softmax normalization:
\begin{equation}
\label{inner product}
    p_l(c^{i(t)}_l) = \text{softmax}({h^{i(t)}_l}^\top \text{emb}(c^{i(t)}_l)),
\end{equation}
where $\text{emb}(c^{i(t)}_l)$ denotes the vector representation of the ground-truth codeword at level $l$. Therefore, the final score for item $i(t)$ is defined as aggregating the log-probabilities across all layers:
\begin{equation}
    \pi(i(t)) = \log \prod_{l=1}^{m}p_l(c^{i(t)}_l) = \sum_{l=1}^{m} \log p_l(c^{i(t)}_l).
\label{eq: IAP_item}
\end{equation}

\subsubsection{Listwise direct preference optimization.}
Although user preference alignment~\cite{liu2025reinforcement,wu2024beta} has been successfully applied to existing methods~\cite{deng2025onerec,chen2025onesearch,zhou2025onerec}, we argue that they commonly rely on training a reward model to achieve the alignment process, which hinders the possibility of end-to-end training and is unfavorable for iterative optimization in real-world production. Moreover, these approaches typically utilize only pairwise loss to guide learning, limiting their ability to model the prevalent partial order relationships inherent in recommendation systems.

To address these limitations, we propose Listwise Direct Preference Optimization (LDPO), which directly leverages high-quality user feedback data of each session as supervision signals to guide the learning process. As illustrated in Figure~\ref{fig: Method}, we model the partial-order signals at four hierarchical levels $\mathcal{G}_4$, $\mathcal{G}_3$, $\mathcal{G}_2$, and $\mathcal{G}_1$: purchase, click, exposure without click, and pseudo-exposure (i.e., items selected by the ranking module but not actually shown to users). During training, LDPO requires that the scores for items within one of the four levels are strictly smaller than those of items with a higher level, which is beyond the widely adopted pair-wise loss.

Formally, given the historical sequence $x^j$ with length $T$ for user $j$ in Equation~\ref{eq:autoregressive_decoding}, we modified it as follows to enable the efficient calculation towards different preferences in a single sample:
\begin{equation}
x^j_{\text{train}}=[\underbrace{c^{i(1)}_1, \dots, c^{i(1)}_m, \dots, c^{i(T)}_1, \dots, c^{i(T)}_m}_{\text{User History}},\underbrace{c^{p_1}_1, \dots, c^{p_1}_m,\dots,c^{p_K}_m}_{\text{Target Items}}],
\label{eq:sequence}
\end{equation}
where $p_k$ denote the items from different levels $\mathcal{G}_k$. Each target item of different hierarchies is modelled with respect to the historical interactions without any mutual interference. Therefore, to ensure that the model does not attend to other target items of the same session, the attention mask is modified as:
\begin{equation}
    M_{t_it_j} = 
    \begin{cases} 
    0, & \text{if } t_j \le t_i \;\land\; ( t_j \in x^j \lor \exists k, \{t_i, t_j\} \subseteq p_k ) \\
    -\infty, & \text{otherwise}
    \end{cases}.
\end{equation}
The mask $M_{t_it_j}$ allows attention from query $t_i$ to key $t_j$ if and only if the key position $j$ precedes or equals the query position $i$ ($j \le i$), and the key $t_j$ either belongs to the historical sequence or belongs to the same target item as the query $t_i$.

The LDPO loss is then applied over items from all four preference levels $ \mathcal{G}_1, \dots, \mathcal{G}_4 $, with the following formulation: 
\begin{equation}
\mathcal{L}_{\text{LDPO}} = -\sum_{j=1}^{3} \sum_{k \in \mathcal{G}_{j+1}} \log \left( \frac{e^{\beta \pi(k)}}{e^{\beta \pi(k)} + \sum_{i \in \mathcal{G}_1 \cup \cdots \cup \mathcal{G}_j} e^{\beta \pi(i)}} \right),
\end{equation}
where $ \pi(k)$ is the IAP score of item $k$ defined in Equation~\ref{eq: IAP_item}.

In addition to the LDPO loss, the sequence of Equation~\ref{eq:sequence} can also be used for the NTP loss defined in Equation~\ref{eq:autoregressive_decoding} with minor modifications. 
Specifically, we broaden the definition of positive labels to encompass purchase, click, exposure without click, and pseudo-exposure:
\begin{equation}
\label{positive}
\mathcal{G}_{\text{pos}} = \{\mathcal{G}_1,\mathcal{G}_2,\mathcal{G}_3, \mathcal{G}_4\} .
\end{equation}
The NTP loss is then computed as:
\begin{equation}
    \mathcal{L}_{\text{NTP}} = \frac{1}{N_u} \sum_{j=1}^{N_u} \sum_{p \in \mathcal{G}_{\text{pos}}} \sum_{l=1}^{m} -\log P_{\theta, \tau}(c^{p}_l \mid x^j, c^{p}_{<l}).
\end{equation}
To prevent GR from the overemphasis on preference alignment, we train the NTP and LDPO objectives in a unified manner:
\begin{equation}
\label{IAP_loss}
\mathcal{L}_{\text{IAP}} =  \mathcal{L}_{\text{NTP}} + \alpha \cdot \mathcal{L}_{\text{LDPO}} ,
\end{equation}
with $ \alpha $ being a hyperparameter. Another benefit is that the co-optimization could leverage the full set of user interactions, rather than relying on a small subset of users to approximate global behavior. This prevents the model from learning spurious patterns or biases that may be specific to certain subpopulations and do not generalize well to broader user scenarios~\cite{zhou2025onerec,chen2025onesearch}.

\subsection{Refined Scoring Phase}
\label{RSP}

Despite the effectiveness to model the partial order of user preferences, IAP still relies on an \textbf{approximation} through the inner product between each codeword and the hidden state $\{h^{i(t)}_l\}_{l=1}^m$ in Equation~\ref{inner product}. To overcome this limitation, we introduce the Refined Scoring Phase (RSP) with a lightweight module to refine the initial scores by enabling deep interaction between each candidate and the rich semantic information in the input sequence.

\subsubsection{Candidate-Centric Interaction}

Specifically, given the input sequence $\{x^j_{<T}, c^{i(T)}_{<l}\}$ as input, we build upon Equation~\ref{hiden_status} to acquire the next hidden state $h_l^{i(T)}$, where $i(T)$ here denotes the $T$-th item in the sequence and corresponds to the target item. Unlike the standard inner product approach in IAP and vanilla GRs, in Figure~\ref{fig: Method}(c) we further exploit the multi-level preference signals introduced by IAP and retain the Top-$\lambda_l$ codewords at level $l$ as $\mathcal{C}_{l}^{i(T)}$:
\begin{equation}
    \mathcal{C}_{l}^{i(T)} = \operatorname{Top-K}\left( \operatorname{Softmax}(h_l^{i(T)}, \tau), K=\lambda_l \right), \forall l\in [1,m]
\end{equation}
where the $\{\lambda_l\}_{l=1}^{m}$ are hyperparameters to control the number of codewords preserved in each level. The codeword set $\mathcal{C}_{l}^{i(T)}$ is then used to perform a comprehensive interaction between these candidates and all previously generated hidden states to refine the estimation in IAP. 
To prevent the information loss of current hidden state $h_l^{i(T)}$, we would update it into $ \mathbf{H} $ in advance:
\begin{equation}
    \mathbf{H} = \left\{ h_2^{i(0)}, \dots, h_1^{i(T-1)}, \dots, h_{l-1}^{i(T)}, h_l^{i(T)} \right\}.
\end{equation}
Subsequently, the \textbf{rank head} made of a \textbf{target attention} operation is performed over $ \mathbf{H} $, treating it as the set of key-value pairs $ (K, V) $, while using each codeword $ \tilde{c}_l \in  \mathcal{C}_{l}^{i(T)}$ itself as the query vector.

Formally, the attention output with full interaction between candidate codeword and input sequences is computed as:
\begin{equation}
    \tilde{h}_{l,\tilde{c}_l}^{i(T)} = \mathrm{Attention}(Q = \text{emb}(\tilde{c}_l), K = \mathbf{H}, V = \mathbf{H}), \forall \tilde{c}_l \in  \mathcal{C}_{l}^{i(T)}.
\end{equation}
The resulting vector $ \tilde{h}_{l,\tilde{c}_l}^{i(T)} $ is then passed through a multi-layer perceptron (MLP) followed by a sigmoid function $\sigma$ to produce a score indicating the relevance of the codeword at this timestamp:
\begin{equation}
    s_{l,\tilde{c}_l}^T = \sigma(\mathrm{MLP}(\tilde{h}_{l,\tilde{c}_l}^{i(T)}))\in (0,1) , \forall \tilde{c}_l \in  \mathcal{C}_{l}^{i(T)},
\end{equation}
where $ \sigma $ denotes the sigmoid function. This procedure is repeated $ m $ times to obtain scores for all $ m $ levels using the corresponding level-specific hidden state $ h_l^{i(T)} $.

During training, we employ the binary cross-entropy (BCE) loss to train the model:
\begin{equation}
    \mathcal{L}_{\text{BCE}} = -\frac{1}{N_u} \sum_{j=1}^{N_u} 
    \sum_{l=1}^{m}
    \sum_{\tilde{c}_l \in  \mathcal{C}_{l}^{i(T)}} 
    y_{l,\tilde{c}_l}^T\log(s_{l,\tilde{c}_l}^T) + (1-y_{l,\tilde{c}_l}^T)\log(1-s_{l,\tilde{c}_l}^T),
\end{equation}
where $y_{l, \tilde{c}_l}^T \in \{0, 1\}$ is the ground truth label, which equals 1 if $\tilde{c}_l$ is the true target item at level $l$ based on Equation~\ref{positive}, and 0 otherwise.

To support the continuous updating of GRs in an online production, the IAP and RSP stages will be jointly optimized. The overall loss function is defined as follows:
\begin{equation}
    \mathcal{L}  = \mathcal{L}_{\text{BCE}} + \mathcal{L}_{\text{IAP}}.
\end{equation}

\subsubsection{Inference Procedure}
During inference, the proposed RankGR framework follows a two-stage decoding pipeline. In the first stage, IAP generates candidate codewords at each level $ l \in [1, m] $ and retains the top-$\lambda_l$ candidates based on their IAP scores. These selected codewords are then passed to RSP for further measurement.

At each level $ l $, the top-$\mathcal{B}_l$ codewords with the highest refined scores are retained as partial candidates for beam search, where $\mathcal{B}_l$ denotes the beam size at that level. As illustrated in Figure~\ref{fig: Method}(c), the overall score of an item is computed as the product of the per-level codeword scores:
\begin{equation}
    s^T(\tilde{c}_1, \dots, \tilde{c}_m) = \log \prod_{l=1}^{m} s_{l,\tilde{c}_l}^T
    = \sum_{l=1}^{m} \log s_{l,\tilde{c}_l}^T.
    \label{RSP_score}
\end{equation}

\subsection{System Deployment}
\label{deployment}

\begin{figure}[htb]
    \centering
    \vspace{-0.2cm}
    \includegraphics[width=1.0\linewidth]{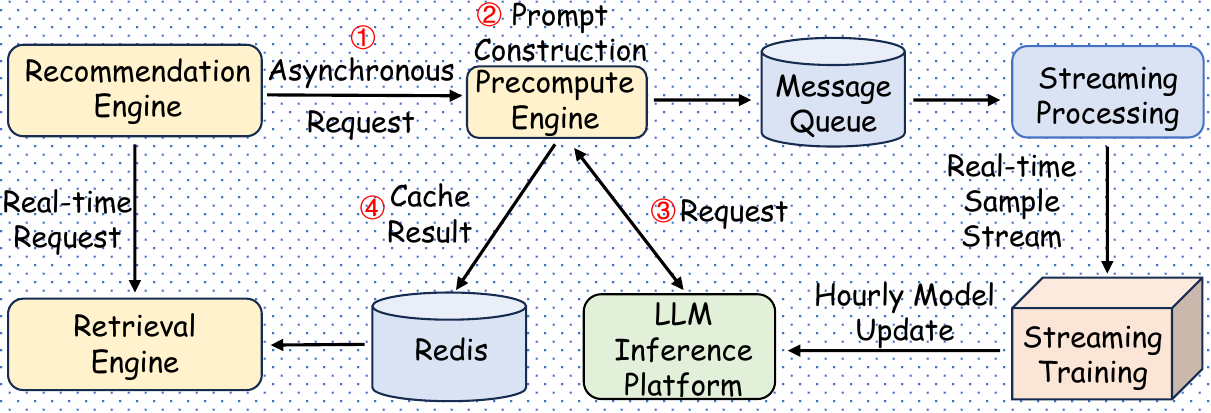}
    \vspace{-0.5cm}
    \caption{Detailed deployment process of RankGR.}
    \label{fig:deployment}
\end{figure}

The detailed deployment architecture of the proposed \textbf{RankGR} framework is illustrated in Figure~\ref{fig:deployment}. To address the inherent latency challenges of LLMs in real-time recommendation scenarios, we design an \textbf{asynchronous} architecture that separates heavy model inference from online serving, while maintaining model freshness through streaming training.

\subsubsection{Asynchronous Pre-computation Architecture}
To reconcile the high inference cost of LLMs with the strict low-latency requirements of online serving, RankGR adopts an asynchronous pre-computation strategy. The core innovation lies in the \textit{latency masking} technique within the workflow:

\begin{itemize}[leftmargin=*]
    \item \textbf{Early Triggering:} The Recommendation Engine initiates the \textit{Asynchronous Request} immediately at the onset of the user request lifecycle. This strategic placement creates a critical time window (typically $>100$ms) where LLM inference runs in parallel with other system tasks (e.g., request parsing, feature extraction).
    
    \item \textbf{Prompt Construction:} Upon receiving the signal, the Precompute Engine constructs context-aware prompts. Since this occurs asynchronously, it allows for sophisticated prompt engineering without blocking the main request thread.
    
    \item \textbf{LLM Inference:} The constructed prompts are dispatched to the LLM Inference Platform. To strictly adhere to the real-time constraints, we optimize the engine using the RTP-LLM framework~\cite{rtp-llm}. Techniques such as load balance, KV cache reuse, and kernel-level fusion are employed to reduce the inference latency to 60ms, ensuring generation within the reserved time window.
    
    \item \textbf{Caching Strategy:} The candidate items generated by IAP and RSP of RankGR are stored in Redis. By the time the main recommendation pipeline reaches the Retrieval Engine, the generated items are already populated in the cache, enabling pseudo-real-time retrieval.
\end{itemize}

\subsubsection{Real-time Retrieval and Serving}
For online requests, the system prioritizes responsiveness. The Recommendation Engine issues a \textit{Real-time Request} to the Retrieval Engine. Instead of invoking RankGR directly—which would incur unacceptable delays—the Retrieval Engine fetches the pre-computed results from Redis. This design effectively decouples the inference latency from the query-time latency, ensuring the system remains responsive even under high concurrency.

\subsubsection{Streaming Training and Model Updates}
To capture shifting user preferences and evolving data distributions, RankGR incorporates a closed-loop online learning mechanism:

\begin{itemize}[leftmargin=*]
    \item \textbf{Data Ingestion:} Real-time interaction data flows into a Message Queue and is processed to generate a \textit{Real-time Sample Stream}.
    
    \item \textbf{Online Learning:} These samples feed into the Streaming Training module, which performs incremental updates of RankGR.
    
    \item \textbf{Model Synchronization:} The system executes an \textit{Hourly Model Update}, synchronizing the fine-tuned weights back to the LLM Inference Platform. This ensures that the deployment remains robust and adaptive to temporal distribution shifts.
\end{itemize}

\section{Experiment}

In this section, we conduct a series of offline experiments on both research and industrial datasets, as well as an online A/B test in the "Guess You Like" section of Taobao's homepage recommendation to demonstrate the effectiveness of the proposed RankGR.
\begin{table}[ht]
\tabcolsep=4.5pt
\vspace{-0.0cm}
\caption{Statistics of the dataset.}
\vspace{-0.3cm}
\label{tab:dataset}
\begin{tabular}{@{}ccccc@{}}
\toprule & \textbf{\#users} & \textbf{\#Items} & \textbf{\#Interactions} & \textbf{Sparsity} \\ \midrule
\textit{Amazon} & 2,524,981 & 714,957 & 12,530,806               & 99.98\%        \\
\textit{Taobao}  &   0.2 billion  &  1.0 billion & 0.9 trillion  & 99.99\%           \\ 
\bottomrule
\end{tabular}
\tabcolsep=7pt
\vspace{-0.5cm}
\end{table}
\subsection{Experimental Setting}

\subsubsection{Dataset}

We evaluate RankGR on two datasets of varying sizes, with their statistical characteristics summarized in Table~\ref{tab:dataset}. Amazon Clothing\_Shoes\_and\_Jewelry\footnote{https://amazon-reviews-2023.github.io/} contains user reviews collected from Amazon in 2023 by McAuley Lab, where each user-item interaction contains a user-id, an item-id, and the corresponding timestamp~\cite{hou2024bridging}. The Taobao dataset is derived from industrial scenarios on Taobao. We randomly sampled data from traffic logs of Taobao from January 11th to January 18th, 2026, forming the Taobao dataset, which is suitable for large-scale distributed training models and better reflects real-world recommendation tasks.

\subsubsection{Baselines}
To provide a reliable comparison, we select six traditional retrieval-based methods and three generative retrieval-based methods to serve as additional baselines:

\begin{itemize}[leftmargin=*]
    \item \textbf{YouTubeDNN}~\cite{covington2016deep} models user preferences by aggregating historical item embeddings through average pooling, forming a compact user representation for efficient candidate generation.
    \item \textbf{SASRec}~\cite{kang2018self} proposes to model user behavior sequences with the self-attention mechanisms in the Transformer framework, capturing complex dependencies among items effectively.
    \item \textbf{Bert4Rec}~\cite{sun2019bert4rec} extends the BERT pre-training strategy to sequential recommendation by masking and reconstructing items in the sequence, thereby learning rich contextual representations.
    \item \textbf{Caser}~\cite{tang2018personalized} employs convolutional operations over user interactions to capture both local and global patterns, making it flexible to model short-term and long-term interests.
    \item \textbf{NextItNet}~\cite{yuan2019simple} utilizes dilated convolutions to model long-range dependencies in sequence, providing an efficient alternative to recurrent-based architectures for next-item prediction.
    \item \textbf{CORE}~\cite{hou2022core} introduces an encoder that generates session representations via a linear weighted sum of item embeddings, ensuring consistency between session and item representations within the same latent space.
    \item \textbf{HSTU}~\cite{zhai2024actions} reformulates the standard retrieval task as a sequential transduction problem, enabling systematic use of feature redundancies and enhancing overall efficiency.
    \item \textbf{TIGER}~\cite{rajput2023recommender} pioneers generative retrieval in recommendation that leverages semantic identifiers and sequence-to-sequence models to predict the next item a user might interact with.
    \item \textbf{FORGE}~\cite{fu2025forge} optimizes the generation of semantic identifiers from the perspective of collaborative information and ID collisions, thus preserving the effectiveness and utilization of SIDs.
\end{itemize}

\subsubsection{Implementation Details}
The SID generation in RankGR is implemented based on the recently proposed FORGE framework~\cite{fu2025forge}. For generative retrieval tasks on both the Amazon and Taobao datasets, we utilize the pretrained Qwen2.5-0.5B-Instruct~\cite{qwen2.5} as the base model. The historical user behaviors are truncated to a maximum length of 2000 tokens. The model is fine-tuned on the respective dataset with a maximum token limit of 2304, where the source and target sequence lengths are set to 2048 and 256, respectively. For the Amazon and Taobao datasets, the training is carried out for five and one epochs, respectively, using a per-device batch size of 40 with 32 in-house PPU-ZW810E GPUs. A linear learning rate scheduler is employed with a base learning rate of $5 \times 10^{-5}$, and all training procedures are conducted in bfloat16 precision to improve computational efficiency. During inference, we apply the dynamic beam search to the two-level SIDs, where the beam sizes are configured as $\{500, 1400\}$. Notably, for the RSP stage, we fix the hyperparameter $\{\lambda_l\}_{l=1}^{2}$ to $\{1400, 1400\}$ to perform more rigorous Top-$\lambda$ selection.

\begin{table*}[htb]
    \centering
    \caption{The performance of evaluated methods on the Taobao dataset.}
    \label{table:taobao_result}
    \vspace{-0.3cm}
    \renewcommand{\arraystretch}{1.1}
    \setlength{\tabcolsep}{2.5pt} 
    \resizebox{0.8\textwidth}{!}{
    \begin{tabular}{c|cccccccccc}
        \toprule[1.5pt]
        \multirow{2}{*}{\textbf{Method}} &
        \multicolumn{5}{c}{\textbf{Click}} & 
        \multicolumn{5}{c}{\textbf{PV}} \\
        \cmidrule(lr){2-6} \cmidrule(lr){7-11}
        & \textbf{HR@20} & \textbf{HR@100} & \textbf{HR@500} & \textbf{HR@1000} & \textbf{HR@2000} &
        \textbf{HR@20} & \textbf{HR@100} & \textbf{HR@500} & \textbf{HR@1000} & \textbf{HR@2000} \\
        \midrule
        \midrule
        \multicolumn{11}{c}{\cellcolor{orange!10}\textcolor{blue}{\textit{Traditional Retrieval Methods}}} \\
        \cline{1-11}
        YoutubeDNN & 2.46\% & 5.36\% & 9.28\% & 11.01\%	 & 12.62\% & 0.28\% & 1.91\% & 5.29\% & 6.77\% & 7.49\% \\
        SASRec & 3.72\% & 8.34\% & 13.01\% & 15.72\% & 18.18\% & 0.40\% & 2.02\% & 7.42\% & 12.40\% & 16.40\% \\
        Bert4Rec & 3.73\% & 8.38\% & 13.12\% & 15.80\% & 18.35\% & 0.31\% & 2.13\% & 7.95\% & 12.96\% & 16.75\% \\
        Caser & 3.17\% & 7.96\% & 11.80\% & 14.67\% & 16.88\% & 0.29\% & 1.89\% & 6.98\% & 11.73\% & 15.42\% \\
        NextItNet & 3.58\% & 8.16\% & 12.64\% & 15.44\% & 17.82\% & 0.32\% & 1.97\% & 7.10\% & 11.90\% & 15.80\% \\
        CORE & 4.20\% & 9.18\% & 15.90\% & 18.85\% & 21.60\% & 0.39\% & 2.34\% & 8.10\% & 13.20\% & 17.40\%\\
        \cline{1-11}
        \multicolumn{11}{c}{\textcolor{purple}{\cellcolor{orange!10}\textit{Generative Retrieval Methods}}} \\
        \cline{1-11}
        HSTU & 4.28\% & 9.26\% & 16.10\% & 19.00\% & 21.80\% & 0.57\% & 2.64\% & 9.04\% & 14.25\% & 19.67\% \\
        TIGER & 7.57\% & 16.51\% & 28.61\% & 33.91\% & 38.88\% & 1.81\% & 6.40\% & 17.85\% & 23.89\% & 31.31\% \\
        FORGE & 9.54\% & 20.59\% & 36.98\% & 43.92\% & 50.19\% & 2.37\% & 7.96\% & 20.66\% & 28.10\% & 35.66\% \\
        \rowcolor{lightblue}
        RankGR & \textbf{12.28\%} & \textbf{26.79\%} & \textbf{46.42\%} & \textbf{55.03\%} & \textbf{63.10\%} & \textbf{2.87\%} & \textbf{10.11\%} & \textbf{26.53\%} & \textbf{36.07\%} & \textbf{45.83\%} \\
        \bottomrule[1.5pt]
    \end{tabular}
    }
    \vspace{-0.3cm}
\end{table*}

\subsubsection{Evaluation Metrics}
Following prior work, we adopt the Hit Rate (HR) as the primary evaluation metric to assess the performance of the retrieval model~\cite{fu2025forge,wang2026pi2i,meng2025user}, which is formulated as:
\begin{equation}
\label{HR}
\text { HR@K }=\frac{1}{|\mathcal{U}|} \sum_{j=1}^{|\mathcal{U}|}\frac{I^j_{K} \cap I^j_{\text{truth}}}{I^j_{\text{truth}}},
\end{equation}
where $I^j_{K}$ and $I^j_{\text{truth}}$ refer to the top-K retrieved items and the true interactions, respectively.
To further evaluate the capability of capturing hierarchical user intents, the experiments on the Taobao dataset typically report hit rates over different levels: click and page view (PV). These metrics are denoted as HR\_Click and HR\_PV, respectively. For brevity, we omit the common prefix and refer to them as \textbf{Click} and \textbf{PV} throughout the paper.

\begin{table}[hbt]
    \centering
    \vspace{-0.15cm}
    \caption{The performance of evaluated methods on the Amazon dataset.}
    \label{table:method_result}
    \vspace{-0.3cm}
    \renewcommand{\arraystretch}{1.1}
    \resizebox{0.8\linewidth}{!}{
    \begin{tabular}{lccccc}
        \toprule[1.5pt]
        \textbf{Method} & \textbf{HR@20} & \textbf{HR@50} & \textbf{HR@100}  & \textbf{HR@500} \\
        \midrule
        \midrule
        \multicolumn{5}{c}{\textcolor{blue}{\cellcolor{orange!10}\textit{Traditional Retrieval Methods}}} \\
        \midrule
        YoutubeDNN & 1.74\% & 2.68\% & 3.85\% & 8.56\%  \\ 
        SASRec & 1.81\% & 2.82\% & 4.09\% & 9.22\% \\ 
        Bert4Rec & 1.80\% & 2.84\% & 4.10\% & 9.23\%  \\  
        Caser & 1.73\% & 2.65\% & 3.81\% & 8.50\% \\ 
        NextItNet & 1.71\% & 2.64\% & 3.78\% & 8.51\% \\ 
        CORE & 1.47\% & 2.67\% & 3.85\% & 9.29\% \\
        \midrule
        \multicolumn{5}{c}{{\textcolor{purple}{\cellcolor{orange!10}\textit{Generative Retrieval Methods}}}} \\
        \midrule
         HSTU & 1.82\% & 2.85\% & 4.16\% & 9.43\% \\
        TIGER & 1.02\% & 1.98\% & 2.47\% & 5.71\% \\ 
        FORGE & 2.12\% & 3.04\% & 4.56\% & 9.79\% \\ 
        \rowcolor{lightblue}
        RankGR & \textbf{2.27\%} & \textbf{3.13\%} & \textbf{4.68\%} & \textbf{9.96\%} \\
        \bottomrule[1.5pt]
    \end{tabular}
    }
    \vspace{-0.4cm}
\end{table}

\subsection{Overall Performance}

In this section, we evaluate the effectiveness of RankGR on both the Taobao and Amazon datasets through the comparison with retrieval baselines. The detailed results are summarized in Table~\ref{table:taobao_result} and Table~\ref{table:method_result}, from which we can conclude the following observations: \textbf{\romannumeral1)} \textbf{More sophisticated neural architectures could lead to better performance.} Among traditional retrieval models, YouTubeDNN achieves significantly lower performance across all metrics, which can be attributed to its relatively simple model structure that fails to capture sequential user patterns effectively. In contrast, more advanced models like SASRec, Bert4Rec, and CORE demonstrate better performance by leveraging sequence modeling and attention mechanisms. \textbf{\romannumeral2)} \textbf{Semantic identifiers do benefit the retrieval task compared with the single token identifiers.} On the Taobao dataset, the semantic identifier-based approaches TIGER and FORGE achieve superior results compared to HSTU with single-token item identifiers. This indicates the advantage of using semantic identifiers in modeling fine-grained item-related interactions. However, on the Amazon dataset, TIGER underperforms HSTU, which we attribute to potential noise or inconsistency in the multi-modal information associated with items in Amazon. In contrast, FORGE performs well on both datasets, indicating that further optimization of SIDs can effectively mitigate such issues and improve robustness. \textbf{\romannumeral3)} \textbf{RankGR achieves state-of-the-art performance across all evaluation metrics on both datasets.} Compared to existing retrieval methods, RankGR shows significant improvements across all metrics on both datasets. This improvement validates the necessity of our two-phase design: the Initial Assessment Phase enables coarse-grained preference alignment via listwise direct preference optimization, while the Refined Scoring Phase refines the relevance scores through deep interaction between candidates and historical sequences. Notably, on the Amazon dataset, RankGR achieves the most significant improvements in top-$K$ metrics with smaller $K$, which indicates that RSP effectively refines item scores and promotes relevant items to higher ranks. These results demonstrate the capability of RankGR to identify items that users are more likely to interact.

\begin{table}[htb]
    \centering
    \vspace{-0.1cm}
    \caption{Ablation study of different model variants on the Taobao dataset.}
    \label{table:ablation_study}
    \vspace{-0.3cm}
    \renewcommand{\arraystretch}{1.2}
    \setlength{\tabcolsep}{4pt}
    \resizebox{1.0\columnwidth}{!}{
    \begin{tabular}{llccccccc}
        \toprule[1.5pt]
        \multicolumn{2}{c}{\textbf{Metric}} & \textbf{RankGR} & \multicolumn{2}{c}{\textbf{w.o. IAP}} & \multicolumn{2}{c}{\textbf{w.o. RSP}} &  \multicolumn{2}{c}{\textbf{w.o. Both}}  \\
        \midrule
        \midrule
        \multirow{5}{*}{\textbf{Click}} 
            & HR@20     & \textbf{12.28\%} & 10.04\% & {\color{mygreen}\small -18.28\%} & 10.99\% & {\color{mygreen}\small -10.44\%} & 9.54\% & {\color{mygreen}\small -22.34\%} \\
            & HR@100    & \textbf{26.79\%} & 21.44\% & {\color{mygreen}\small -19.96\%} & 24.72\% & {\color{mygreen}\small -7.72\%} & 20.59\% & {\color{mygreen}\small -23.14\%} \\
            & HR@500    & \textbf{46.42\%} & 38.41\% & {\color{mygreen}\small -17.23\%} & 44.36\% & {\color{mygreen}\small -4.42\%} & 36.05\% & {\color{mygreen}\small -20.33\%} \\
            & HR@1000   & \textbf{55.03\%} & 46.00\% & {\color{mygreen}\small -16.41\%} & 52.93\% & {\color{mygreen}\small -3.82\%} & 43.92\% & {\color{mygreen}\small -20.19\%} \\
            & HR@2000   & \textbf{63.10\%} & 52.75\% & {\color{mygreen}\small -16.40\%} & 60.95\% & {\color{mygreen}\small -3.40\%} & 50.19\% & {\color{mygreen}\small -20.46\%} \\
        \midrule
        \multirow{5}{*}{\textbf{PV}} 
            & HR@20     & \textbf{2.87\% }& 2.47\% & {\color{mygreen}\small -13.98\%} & 2.62\% & {\color{mygreen}\small -8.71\%} & 2.37\% & {\color{mygreen}\small -17.42\%} \\
            & HR@100    & \textbf{10.11\%} & 8.46\% & {\color{mygreen}\small -16.32\%} & 9.24\% & {\color{mygreen}\small -8.65\%} & 7.96\% & {\color{mygreen}\small -21.23\%} \\
            & HR@500    & \textbf{26.53\%} & 21.83\% & {\color{mygreen}\small -17.72\%} & 24.41\% & {\color{mygreen}\small -7.99\%} & 20.66\% & {\color{mygreen}\small -22.11\%} \\
            & HR@1000   & \textbf{36.07\%} & 29.62\% & {\color{mygreen}\small -18.10\%} & 33.57\% & {\color{mygreen}\small -6.95\%} & 28.10\% & {\color{mygreen}\small -22.10\%} \\
            & HR@2000   & \textbf{45.83\%} & 37.46\% & {\color{mygreen}\small -18.26\%} & 43.19\% & {\color{mygreen}\small -5.75\%} & 35.66\% & {\color{mygreen}\small -22.19\%} \\
        \bottomrule[1.5pt]
    \end{tabular}
    }
    \vspace{-0.2cm}
\end{table}

\subsection{Ablation Study}
\label{Ablation Study}

To further assess the contribution of each component in RankGR, we conduct ablation experiments on the Taobao dataset by separately removing IAP, RSP, or both. The results are summarized in Table~\ref{table:ablation_study}, from which our key observations are as follows: 

\textbf{\romannumeral1) w/o IAP} removes the listwise preference modeling. Despite retaining codeword-level autoregressive decoding and the refined assesment in RSP, the performance drops significantly across all metrics. For example, HR@20 for Click decreases by 18.28\%, indicating that hierarchical user preferences play a significant role in the usage of GRs in recommendation. 

\textbf{\romannumeral2) w/o RSP} disables the deep interaction between candidate items and historical sequences. While this variant could benefit from the partial-order modelling of IAP, lacking the inherent interactions between the input sequence and candidate codewords in RSP could still introduce substantial performance degradation. As we discovered in Table~\ref{table:taobao_result}, the performance degradation becomes more severe as $ K $ decreases, providing further evidence that RSP indeed promotes interesting items to higher ranks with the scoring module.

\textbf{\romannumeral3) w/o Both} completely removes both IAP and RSP, reverting to a basic GR model. This leads to more than 20\% performance degradation, demonstrating the complementary roles of IAP and RSP in achieving strong recommendation performance.

\begin{figure}[h]
  \centering
  \vspace{-0.15cm}
  \includegraphics[width=\linewidth]{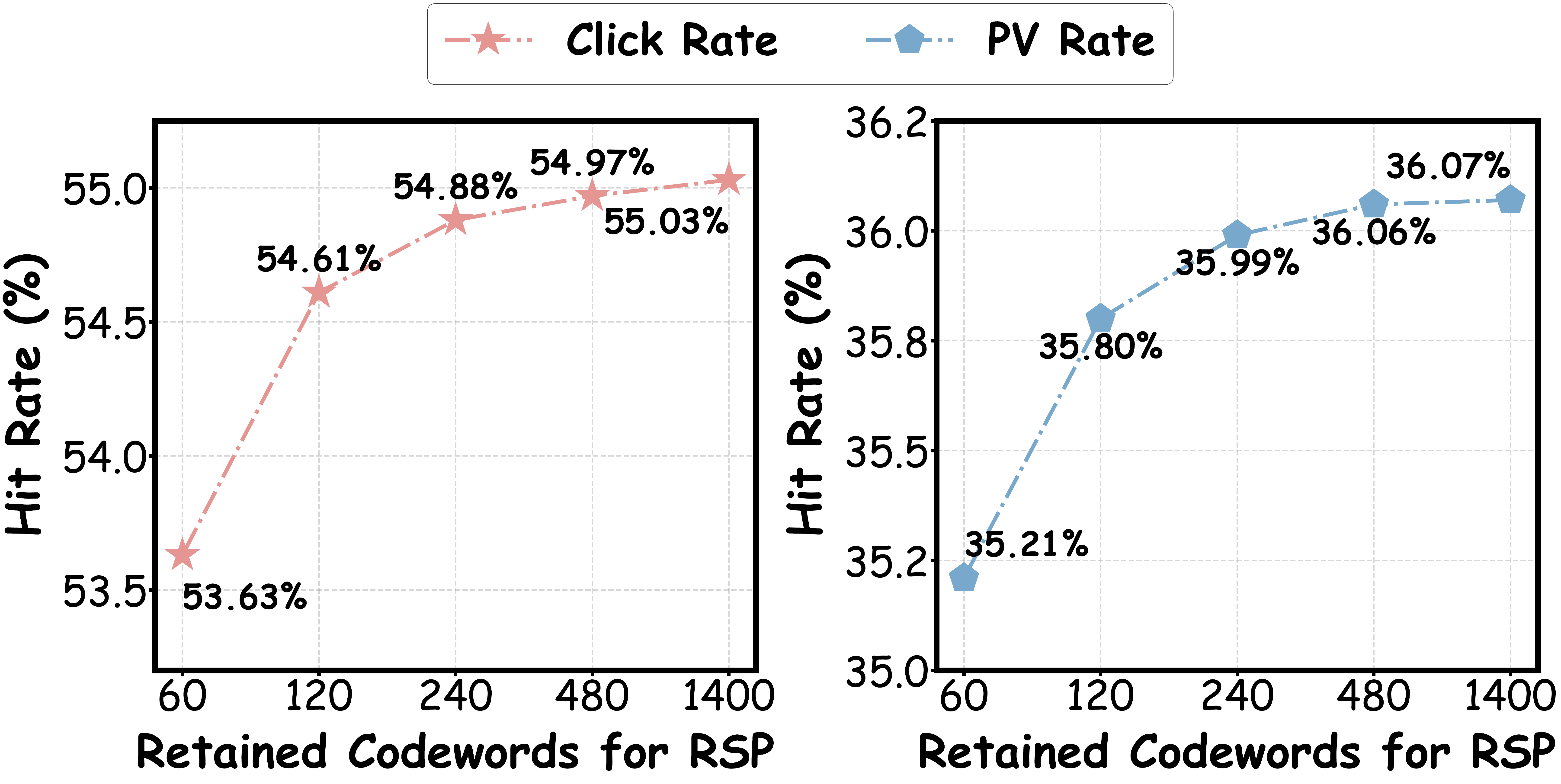}
  \vspace{-0.7cm}
  \caption{Performance of RankGR in terms of HR@1000 with respect to the number of retaining codewords for RSP.}
  \label{fig:RSP_codeword}
  \vspace{-0.5cm}
\end{figure}

\subsection{In-depth Analysis}

\subsubsection{Analysis of the number of retaining codewords in RSP}

We evaluate the impact of different numbers of top codewords ($\lambda_2$) retained from IAP during RSP. The analysis of $\lambda_1$ for the first codeword is omitted, as preliminary experiments showed that its value has little impact on RSP and it typically represents broad categories (e.g., electronics, clothing). As shown in Figure~\ref{fig:RSP_codeword}, both Click and PV rates increase once we increase $\lambda_2$, indicating that a larger candidate space is beneficial for capturing diverse and relevant items. In contrast, the performance degradation at lower $\lambda_2$ values (e.g., 60) suggests that preserving too few candidates from IAP may lead to premature pruning of potentially relevant items, demonstrating the limitations of IAP. This result highlights the importance of maintaining appropriately codewords from IAP for RSP to balance accuracy and efficiency.

\begin{figure}[htb]
  \centering
  \includegraphics[width=\linewidth]{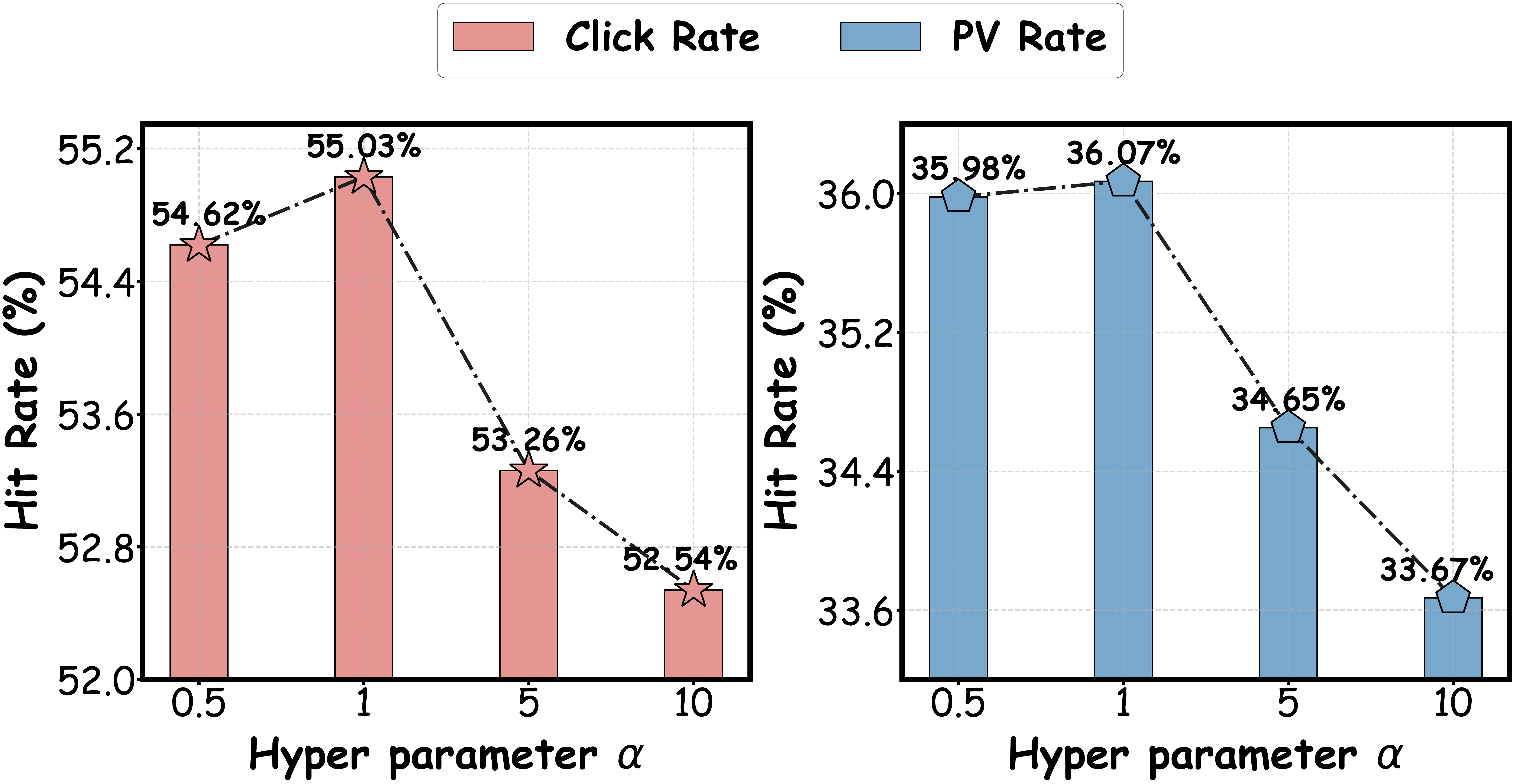}
  \vspace{-0.65cm}
  \caption{Performance of RankGR  in terms of HR@1000 on the Taobao dataset with respect to the hyperparameter $\alpha$.}
  \label{fig:IAP_ALPHA}
  \vspace{-0.55cm}
\end{figure}

\subsubsection{Analysis of the hyperparameter for the NTP and LDPO loss}

To investigate how the hyperparameter $\alpha$  affects the training dynamics in IAP and balances the contributions of listwise preference optimization $\mathcal{L}_{\text{LDPO}}$ and next-token prediction $\mathcal{L}_{\text{NTP}}$, we vary the corresponding value of $\alpha$ and plot the impact in Figure~\ref{fig:IAP_ALPHA}.  
As $\alpha$ increases from 0.5 to 1, both Click and PV rates improve, showing that a stronger emphasis on LDPO enhances the ability of GRs to capture hierarchical user preferences. However, when $\alpha$ exceeds 1, the performance starts to decline. We attribute this phenomenon to the overemphasis on preference alignment that undermines the basic autoregressive generation objective.

\begin{table}[htb]
    \centering
    \vspace{-0.25cm}
    \caption{Online experiments from 20st Jan to 27th Jan.}
    \vspace{-0.3cm}
    \label{tab:online_study}
    \renewcommand{\arraystretch}{1.1} 
    \setlength{\tabcolsep}{3.5pt} 
    \resizebox{0.72\columnwidth}{!}{
        \begin{tabular}{lccc}
            \toprule[1.5pt]
            \textbf{Model} & \textbf{PVR} & \textbf{IPV} & \textbf{Transaction Count} \\
            \midrule
            \midrule
            Base & 36.79\% & - & - \\
            RankGR & \textbf{49.88\%} & \textbf{+1.08\%} & \textbf{+0.57\%} \\
            \bottomrule[1.5pt]
        \end{tabular}%
    }
    \vspace{-0.5cm}
\end{table}

\subsection{Online Study}

We conduct online A/B testing of RankGR in the "Guess You Like" section on Taobao for the homepage recommendation, one of the most prominent recommendation scenarios of e-commerce in China. This deployment replaces the existing GR model (i.e., GRs without IAP and RSP) with our RankGR. 

The results from January 20th to January 27th are shown in Table~\ref{tab:online_study}. With a traffic coverage of 5\%, RankGR brings significant improvements over the online production retrieval-ranking framework across multiple key online metrics. Specifically, with the asynchronous pre-computation architecture and streaming training proposed in Section~\ref{deployment}, RankGR captures 49.88\% of total exposures, indicating that it successfully retrieves about half of the exposed items among all deployed retrieval methods. Additionally, Item Page Views (IPV) and Transaction Count are improved by +1.08\% and +0.57\%, respectively. These improvements consistently confirm the effectiveness of RankGR in enhancing user engagement and commercial performance in real-world industrial settings.

\section{Conclusion}
In this paper, we present RankGR, a novel method that leverages listwise direct preference optimization and an additional lightweight rank head to enhance the training of GRs towards the hierarchy of user interests and in-depth modeling between candidate semantic identifiers and user sequences. Extensive comparisons with other methods and subsequent in-depth analysis on several real-world datasets demonstrate the effectiveness of RankGR. Notably, we further validate its effectiveness by deploying it in the "Guess You Like" section of Taobao and observing the commercial advantages it brings.


\bibliographystyle{ACM-Reference-Format}
\bibliography{sample-base}










\end{document}